# Dispersion-Aware Modeling Framework for Parallel Optical Computing

Ziqi Wei[1,2,3,†,*], Yuanjian Wan[4,†], Yuhu Cheng[1,4], Xiao Yu[4,*], Peng Xie[1,4,*]

[1]Center of Materials Science and Optoelectronics Engineering, University of Chinese Academy of Sciences, Beijing 100049, China
[2]School of physics, Peking University, Beijing 100871, China
[3]School of Engineering and Applied Sciences, Harvard University, Cambridge, MA 02138, USA
[4]Wangzhijiang Innovation Center for Laser, Aerospace Laser Technology and System Department, Shanghai Institute of Optics and Fine Mechanics, Chinese Academy of Sciences, Shanghai 201800, China

[†]These authors contributed equally to this work

**ABSTRACT**. Optical computing represents a groundbreaking technology that leverages the unique properties of photons, with innate parallelism standing as its most compelling advantage. Parallel optical computing like cascaded Mach-Zehnder interferometers (MZIs) based offers powerful computational capabilities but also introduces new challenges, particularly concerning dispersion due to the introduction of new frequencies. In this work, we extend existing theories of cascaded MZI systems to develop a generalized model tailored for wavelength-multiplexed parallel optical computing. Our comprehensive model incorporates component dispersion characteristics into a wavelength-dependent transfer matrix framework and is experimentally validated. We propose a computationally efficient compensation strategy that reduces global dispersion error within a 40 nm range from 0.22 to 0.039 using edge-spectrum calibration. This work establishes a fundamental framework for dispersion-aware model and error correction in MZI-based parallel optical computing chips, advancing the reliability of multi-wavelength photonic processors.

## I. INTRODUCTION.

The relentless growth of large language models and embodied artificial intelligence is pushing the limits of traditional computing, primarily due to the inherent bottlenecks in power efficiency and data transfer bandwidth. In this context, optical computing has re-emerged as a compelling alternative, promising to harness the fundamental properties of light—such as its ultra-high speed and innate parallelism — for transformative computational performance.

The evolution of optical computing has followed several technological pathways, from free-space optics [1–3] to integrated photonic circuits. A practical and widely adopted approach utilizes Mach-Zehnder Interferometer (MZI) meshes [4–7], micro-ring resonators(MRR) [8,9], diffractive photonics [10,11], phase changed materials(PCMs) [12] to perform vector-matrix multiplication optically. The computility of such a system is fundamentally governed by three physical parameters: the modulation frequency, the scale of the programmable matrix, and the number of parallel channels. While significant progress has been made in pushing the frontiers of frequency [13], matrix scale [14], these dimensions are approaching physical ceilings. Besides, optical computing faces inherent challenges as an analog paradigm. To deal with these errors, researchers propose various methods, including calibration procedure [15], propagating auxiliary phases [16], iterative algorithms [4,17].

Wavelength division multiplexing (WDM) emerges as the pivotal strategy to break this scalability barrier. It uniquely exploits the spectral degree of freedom of light, creating multiple parallel computing channels within the same physical hardware. By encoding independent data streams onto different wavelengths, WDM enables a linear scaling of computational throughput without the need for duplicating components or compromising operational speed. In the current stage, microcomb-based multiwavelength source enables over 100 parallel channels optical computing, which increases the compulity by 2 orders of magnitude [7,18,19]. This wavelength-multiplexed parallel optical computing architecture eliminates electronic bottlenecks, accelerating AI tasks and complex simulations while maintaining energy efficiency, paving the way for next-generation computing systems.

*Contact author: ziqiwei@g.harvard.edu; yuxiao@siom.ac.cn; pengx@siom.ac.cn

[†]These authors contributed equally to this work

However, exploiting the wavelength dimension for parallel computing confronts an entirely new class of problems. The critical dependency of the system's response on wavelength introduces intrinsic constraints and complexities—such as spectral non-uniformity and dispersion-induced errors—that have no direct analog in traditional optical computing and must be addressed from first principles. Hereby, we propose a modified dispersive model to bridge this theoretical-practical gap. This work provides a practical and robust framework for MZI-based systems, featuring a dedicated method for the correction of dispersion-induced errors. Furthermore, the underlying approach is generalizable, offering a framework for enhancing accuracy in other parallel optical computing architectures.

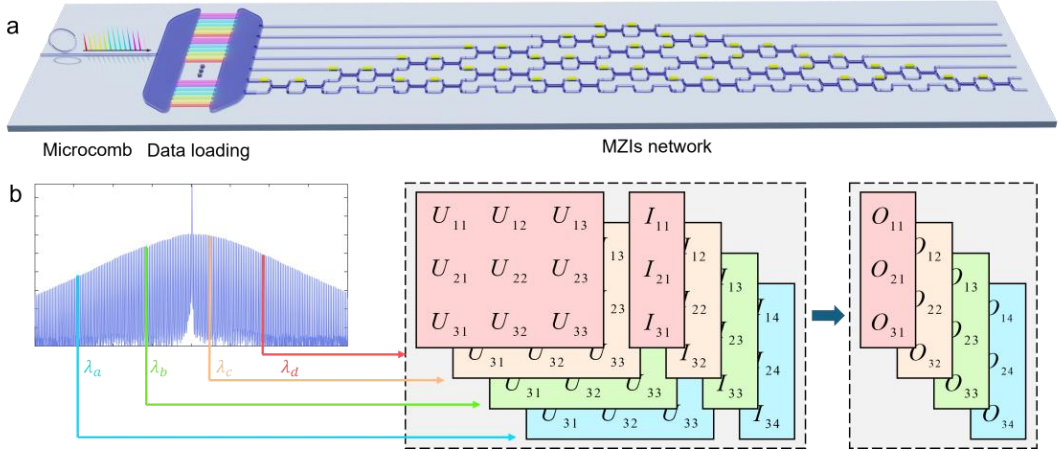

**Fig1. Model of Multi-Wavelength Multiplexing Parallel Optical Computing.** a) Schematic of the Parallel Optical Computing. A multi-wavelength source is utilized, where the wavelength channels are demultiplexed and modulated. Multiple wavelength channels are combined into each channel to facilitate parallel optical computing. b) Spectrum of microcomb multiwavelength light source and its correspondence to dispersive parallel matrix multiplication model. The response of optical computing chips against different wavelengths is modeled as matrices with minor differences.

## II. THEORY

In single wavelength optical computing, matrices can be loaded on MZI mesh by applying [20] and [21] decompositions. However, when transitioning to parallel optical computing, each wavelength channel encoding independent data vectors are multiplexed to the same MZI mesh. The parallel optical computing model is depicted in Fig.1a. The wavelength-dependent variations emerge in the spectral response of individual MZIs. In cascaded interferometric systems, response deviations in individual MZIs propagate as cascading errors, introducing deviations in the programmed matrix and degrading computational consistency as shown in Fig.1b. As no established theoretical framework currently exists to analyze these errors, we propose a comprehensive model to quantify them and develop a correction method.

Generally, these errors can be categorized into two components: dispersion errors and wavelength-dependent transmission loss errors. Transmission losses in photonic circuits primarily comprise propagation loss in straight waveguides and bend loss. In well-fabricated chips, these losses remain relatively uniform across the mesh and exhibit minimal wavelength dependence, as evidenced in subsequent experimental results. With appropriate back-end processing, the impact of this factor is effectively mitigated. Therefore, this work expands the MZI optical computing theory to the multiwavelength regime and concentrates on critical dispersion behavior in a parallel optical computing chip. The dispersion errors predominantly arise in phase shifters, characterized by wavelength-dependent variations of their response function. The dispersion induced by well-designed broadband beam splitter is negligible. The magnitude of these dispersion errors constitutes a critical limitation on both mesh scalability and computing wavelength range. To address this, we develop a theory to systematically quantify these dispersion errors.

To begin with, a single MZI includes an internal phase shifter located between the splitters on one of the modes, an external phase shifter at one output, and two 50-50 beam splitters. The overall operation can be described as a 2×2 matrix $T$:

*Contact author: ziqiwei@g.harvard.edu; yuxiao@siom.ac.cn; pengx@siom.ac.cn

†These authors contributed equally to this work

$$T(\theta,\phi) = \frac{1}{2}\begin{bmatrix} e^{i\phi} & 0 \\ 0 & 1 \end{bmatrix}\begin{bmatrix} 1 & i \\ i & 1 \end{bmatrix}\begin{bmatrix} e^{i\theta} & 0 \\ 0 & 1 \end{bmatrix}\begin{bmatrix} 1 & i \\ i & 1 \end{bmatrix} \quad (1)$$

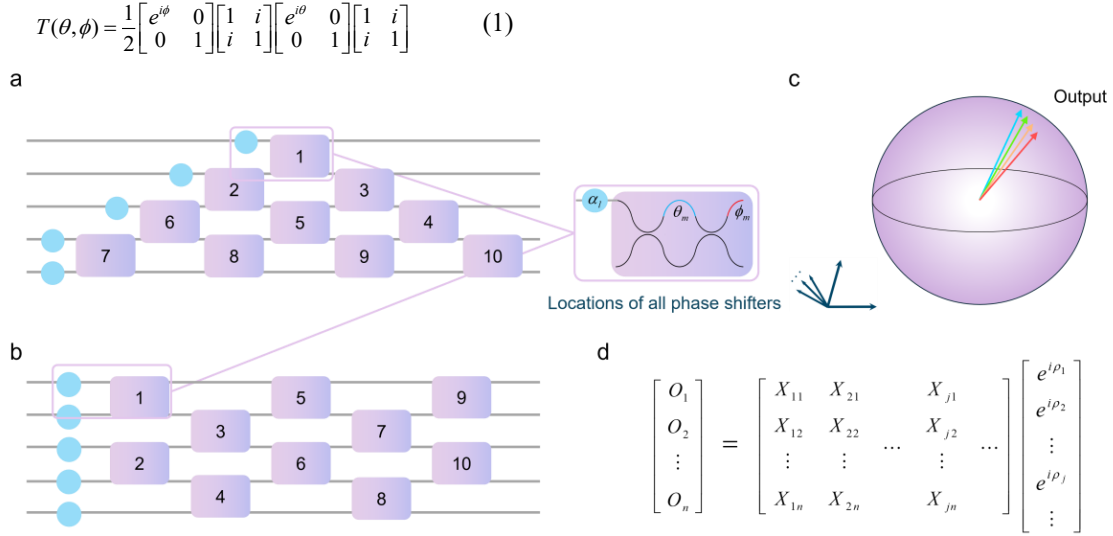

**Fig2. A unique perspective on two MZI mesh computing configurations.** a) A triangle configuration consists of $(n-1)$ rows and $(2n-3)$ columns of MZIs. The blue circle on the $l^{th}$ port represents the input phase $\alpha_l$; the $m^{th}$ violet block represents the $m^{th}$ MZI equipped with two phase shifters characterized by phases $\theta_m, \phi_m$. b) A rectangular configuration consists of $(n-1)$ rows and $n$ columns of MZIs. c) Manifestation of dispersion error as a wavelength-dependent spread of output vectors. A vector-matrix multiplication at different wavelengths results in a distribution of output vectors on a high-dimensional spherical surface. d) The output light vector can be regarded as a constant complex matrix multiplies a phase term vector, where each $\rho_j$ is a linear combination of $n^2$ angles $\theta_m, \phi_m, \alpha_l$.

For MZI mesh, let $I$ and $O$ be the normalized input and output n-dimensional light vector respectively:

$$O = UI, \quad O^\dagger O = I^\dagger I = 1 \quad (2)$$

where $U$ is the overall $n \times n$ unitary matrix of the whole MZI mesh. Let $T_m(\theta_m, \phi_m)$ be $n \times n$ unitary matrix of only the $m^{th}$ MZI (which means $2 \times 2$ matrix $T(\theta_m, \phi_m)$ is the restricted matrix of $T_m(\theta_m, \phi_m)$ on the two related channels), then $U$ is a product of $\left(\frac{1}{2}n(n-1)+1\right)$ matrices:

$$U\left(\theta_1, \cdots, \theta_{\frac{1}{2}n(n-1)}, \phi_1, \cdots, \phi_{\frac{1}{2}n(n-1)}, \alpha_1, \cdots, \alpha_n\right) = \prod_{m=1}^{\frac{1}{2}n(n-1)} T_m(\theta_m, \phi_m) \cdot diag(e^{i\alpha_1}, \cdots, e^{i\alpha_n}) \quad (3)$$

As shown in Fig.2a-c, $\alpha_l$ is the input phase of the $l^{th}$ port, $\theta_m, \phi_m$ represent the angles of two phase shifters in the $m^{th}$ MZI.

For simplicity, let set $\Theta = \{\theta_1, \ldots, \theta_{\frac{1}{2}n(n-1)}\}$, set $\Phi = \{\phi_1, \ldots, \phi_{\frac{1}{2}n(n-1)}\}$. For a given input I, the output O is obtained by the explicit multiplication of all $\left(\frac{1}{2}n(n-1)+1\right)$ matrices as follows (Fig.2d):

$$O\left(\theta_1, \cdots, \theta_{\frac{1}{2}n(n-1)}, \phi_1, \cdots, \phi_{\frac{1}{2}n(n-1)}, \alpha_1, \cdots, \alpha_n\right) = \sum_{j}^{finite} X_j \exp(i\rho_j)$$
$$\rho_j \stackrel{def}{=} \sum_{\theta \in \Theta_j} \theta + \sum_{\phi \in \Phi_j} \phi + \alpha_{a_j} \quad (4)$$

in which $\Theta_j$ is a subset of $\Theta$, $\Phi_j$ is a subset of $\Phi$; $a_j$ is an integer satisfying $1 \le a_j \le n$; $X_j$ is a constant n-dimensional complex vector. The output complex vector is the linear combination of the input. The transformation relation is determined by all $\theta$ and $\phi$ of all possible paths between the input and the output. Notice that for an MZI mesh (triangular or rectangular, see Fig.2a,b), the maximum number of $\theta's$ consisted in $\rho$ equals to the number of MZI columns, while the maximum number of $\phi's$ consisted in $\rho$ equals the number of MZI rows. Thus, we have:

$$\begin{aligned} \text{triangular} &: \quad \max_j(|\Theta_j|) = 2n-3, \quad \max_j(|\Phi_j|) = n-1 \\ \text{rectangular} &: \quad \max_j(|\Theta_j|) = n, \quad \max_j(|\Phi_j|) = n-1 \end{aligned} \quad (5)$$

To perform parallel optical computing, we expect the overall matrix $U$ loaded on the MZI mesh to be a specified value $U^{sp}$. When all $n^2$ angles $\theta_m, \phi_m, \alpha_l$ to be specified values $\theta_m^{sp}, \phi_m^{sp}, \alpha_l^{sp}$, we obtain the desired output $O^{desired} = U^{sp}I$. These angles can be calculated by using deterministic [20,21] or iterative [15,17,22,23] approaches.

*Contact author: ziqiwei@g.harvard.edu; yuxiao@siom.ac.cn; pengx@siom.ac.cn

†These authors contributed equally to this work

However, due to the presence of dispersion, all $n^2$ angles $\theta_m, \phi_m, \alpha_l$ vary with wavelength $\lambda$:

for $m = 1, 2, \ldots, \frac{1}{2}n(n-1)$ and $l = 1, 2, \ldots, n$

$$[\theta_m, \phi_m, \alpha_l, \rho_j, U, O] = [\theta_m, \phi_m, \alpha_l, \rho_j, U, O](\lambda, \lambda_0)$$
$$[\theta_m, \phi_m, \alpha_l, U, O](\lambda_0, \lambda_0) = [\theta_m^{sp}, \phi_m^{sp}, \alpha_l^{sp}, U^{sp}, O^{desired}] \quad (6)$$

in which $\lambda_0$ is the wavelength where the matrix is calibrated.

For simplicity, rewrite the output $O$ as follows:

$$O(\lambda, \lambda_0) = \sum_j^{finite} X_j \exp(i\rho_j(\lambda, \lambda_0)) = \sum_j^{finite} Y_j \exp(i\eta_j(\lambda, \lambda_0)) \quad (7)$$

in which the complex $n$-dimensional complex vector $Y_j$ and the phase variation $\eta_j(\lambda)$ satisfy:

$$Y_j = X_j \exp(i\rho_j(\lambda_0, \lambda_0))$$
$$\eta_j(\lambda, \lambda_0) = \rho_j(\lambda, \lambda_0) - \rho_j(\lambda_0, \lambda_0) \quad (8)$$

The dispersion term of phase $\exp(i\,\eta_j(\lambda, \lambda_0))$ is isolated for subsequent analysis.

Notice that the modulus of the output $\left|O(\theta_1, \ldots, \theta_{\frac{1}{2}n(n-1)}, \phi_1, \ldots, \phi_{\frac{1}{2}n(n-1)}, \alpha_1, \ldots, \alpha_n)\right|$ always equals the modulus of the input $|I|$, hence it is a constant 1. Consequently, for any $t \in \mathbb{R}$, we have:

$$\left|\sum_j^{finite} Y_j \exp(it\eta_j(\lambda, \lambda_0))\right| = 1 \quad (9)$$

which constitutes the prerequisite for the FCM Theorem (see later) and enables us to limit the error of every order.

For three types of angle $\theta$, $\phi$, $\alpha$, assume their limit dispersion errors (Over all possible specified values) are $\Delta_1 \theta(\lambda, \lambda_0), \Delta_1 \phi(\lambda, \lambda_0), \Delta_1 \alpha(\lambda, \lambda_0)$ (See Supplementary Information S2), respectively. The upper limit $\delta_1(\lambda, \lambda_0)$ of all the phase deviations $|\eta_j(\lambda, \lambda_0)|$ can be represented as follows:

$$\max_j \left(|\eta_j(\lambda, \lambda_0)|\right) \leq \delta_1(\lambda, \lambda_0)$$
$$\delta_1(\lambda, \lambda_0) \stackrel{\text{def}}{=} \Delta_1 \theta(\lambda, \lambda_0) \max_j (|\Theta_j|) + \Delta_1 \phi(\lambda, \lambda_0) \max_j (|\Phi_j|) + \Delta_1 \alpha(\lambda, \lambda_0) \quad (10)$$

To rigorously formulate error estimation protocols in the MZI mesh, the following theorem is established as a theoretical foundation:

**FCM (Fourier Conditioned Moment) Theorem:**
Assume a n-dimensional complex-value vector function $Y(\eta)$ on $[-\delta, \delta]$. If for any $t \in \mathbb{R}$, $Y(\eta)$ satisfies:

$$\left|\int_{-\delta}^{\delta} Y(\eta) \cdot e^{it\eta} d\eta\right| = 1 \quad (11)$$

Then

$$\left|\int_{-\delta}^{\delta} Y(\eta) \cdot \eta^k d\eta\right| \leq \delta^k, \forall k \in \mathbb{N}^* \quad (12)$$

Proof: (See Supplementary Information S1 and Fig.3d).

Based on the theorem, the dispersion error of the output can be evaluated as:

$$O(\lambda, \lambda_0) - O^{desired} = \sum_j^{finite} Y_j \left(e^{i\eta_j(\lambda, \lambda_0)} - 1\right) = \sum_{k=1}^{+\infty} \frac{i^k}{k!} \sum_j^{finite} Y_j \eta_j^k \quad (13)$$

Process sums as integrals, apply FCM Theorem and discretize, we can derive an upper bound of the $k^{th}$ order error:

$$\left|\sum_j^{finite} Y_j \eta_j^k\right| \leq \delta_1^k, \ \forall k \in \mathbb{N}^* \quad (14)$$

Finally, one can obtain:

$$\left|O(\lambda, \lambda_0) - O^{desired}\right| \leq e^{\delta_1(\lambda, \lambda_0)} - 1 \quad (15)$$

In practice, since different wavelengths perform parallel optical computing independently, an overall phase term $e^{-i\gamma}$ can be added in the output. Generally, all phase shifters have dispersion in the same direction. Take $\gamma$ as $\pm \frac{\delta_1}{2}$ (same sign as all $\eta_j$), the same process as above yields:

$$\left|e^{-i\gamma} O(\lambda, \lambda_0) - O^{desired}\right| \leq e^{\frac{1}{2}\delta_1(\lambda, \lambda_0)} - 1 \quad (16)$$

which indicates the dispersion error limit of the output.

### III. ERROR CORRECTION STRATEGY AND RESIDUAL ANALYSIS

**A. First order interpolation compensation method**
Guided by the theorem, we introduce a compensation method that employs first order interpolation to counteract dispersion errors, thereby significantly enhancing the spectral consistency of parallel optical computing. For computing wavelengths in range

*Contact author: ziqiwei@g.harvard.edu; yuxiao@siom.ac.cn; pengx@siom.ac.cn

†These authors contributed equally to this work

$[\lambda_{min}, \lambda_{max}]$ with center wavelength $\lambda_0 = \frac{1}{2}(\lambda_{min} + \lambda_{max})$, choose two calibration wavelengths $\lambda_{1,2}$ as follows:

$$\lambda_1 = \lambda_0 - \frac{\lambda_{max} - \lambda_{min}}{2\sqrt{2}}, \quad \lambda_2 = \lambda_0 + \frac{\lambda_{max} - \lambda_{min}}{2\sqrt{2}} \quad (17)$$

The corrected output is presented as follows:

$$O^{corrected} = \frac{(\lambda_2 - \lambda) \cdot O(\lambda, \lambda_1) + (\lambda - \lambda_1) \cdot O(\lambda, \lambda_2)}{\lambda_2 - \lambda_1} \quad (18)$$

The residual error after correction can be expanded into second order and higher order components. We can extract the second order minor or the residual error as follows:

$$O^{desired} - O^{corrected} \cong \frac{(\lambda_2 - \lambda)(\lambda - \lambda_1)}{2(\lambda_0 - \lambda_1)^2} \sum_j^{finite} Y_j \left( \eta_j(\lambda_0, \lambda_1)^2 - i\eta_j(\lambda_0, \lambda_1) - i\eta_j(\lambda_0, \lambda_2) \right) \quad (19)$$

in which the approximately equal sign represents the difference between two sides is at most a third-order minor.

Similar to Eq. (9), for any $t \in \mathbb{R}$, we also have:

$$\left| \sum_j^{finite} Y_j \exp\left( it \left( \eta_j(\lambda_0, \lambda_1) + \eta_j(\lambda_0, \lambda_2) \right) \right) \right| = 1 \quad (20)$$

which constitutes the prerequisite for the FCM Theorem and enables us to limit a part of the second-order error.

For three types of angle $\theta, \phi, \alpha$, their general behaviors regarding input wavelength and calibrated wavelength are shown in Fig.3c. Assume their limit second-order dispersion errors (Over all possible specified values) are $\Delta_2\theta(\lambda_0, \lambda_1), \Delta_2\phi(\lambda_0, \lambda_1), \Delta_2\alpha(\lambda_0, \lambda_1)$ (See Supplementary Information S2), respectively, then we get an upper limit $\delta_2(\lambda_0, \lambda_1)$ of all the second-order phase deviations $\left| \eta_j(\lambda_0, \lambda_1) + \eta_j(\lambda_0, 2\lambda_0 - \lambda_1) \right|$ as follows:

$$\max_j \left( \left| \eta_j(\lambda_0, \lambda_1) + \eta_j(\lambda_0, 2\lambda_0 - \lambda_1) \right| \right) \leq \delta_2(\lambda_0, \lambda_1)$$
$$\delta_2(\lambda_0, \lambda_1) \stackrel{\text{def}}{=} \Delta_2\theta(\lambda_0, \lambda_1) \max_j \left( |\Theta_j| \right) + \Delta_2\phi(\lambda_0, \lambda_1) \max_j \left( |\Phi_j| \right) + \Delta_2\alpha(\lambda_0, \lambda_1) \quad (21)$$

Similar to the previous process, after processing sums as integrals, applying the FCM Theorem and discretizing, we have:

$$\left| \sum_j^{finite} Y_j \eta_j(\lambda_0, \lambda_1)^2 \right| \leq \delta_1(\lambda_0, \lambda_1)^2$$
$$\left| \sum_j^{finite} Y_j \left( \eta_j(\lambda_0, \lambda_1) + \eta_j(\lambda_0, \lambda_2) \right) \right| \leq \delta_2(\lambda_0, \lambda_1) \quad (22)$$

Finally, we derive the limit residual error (the limit of second-order error in the residual error, to be precise) as follows (See Fig.3b):

$$\left| O^{desired} - O^{corrected} \right| \cong \left| \left( O^{desired} - O^{corrected} \right)_2 \right|$$
$$\leq \frac{1}{2} \cdot \left| \frac{(\lambda_2 - \lambda)(\lambda - \lambda_1)}{(\lambda_0 - \lambda_1)^2} \right| \cdot \left( \delta_2(\lambda_0, \lambda_1) + \delta_1(\lambda_0, \lambda_1)^2 \right) \quad (23)$$

As shown in Fig.3b, the residual error peaks at the center and both extremities of the operational wavelength band. Define uniform limit residual error as follows:

$$\delta_*(\lambda_0, \lambda_1) = \frac{1}{2} \left( \delta_2(\lambda_0, \lambda_1) + \delta_1(\lambda_0, \lambda_1)^2 \right) \quad (24)$$

Then for all computing wavelengths, we have:

$$\left| O^{desired} - O^{corrected} \right| \cong \left| \left( O^{desired} - O^{corrected} \right)_2 \right| \leq \delta_*(\lambda_0, \lambda_1) \quad (25)$$

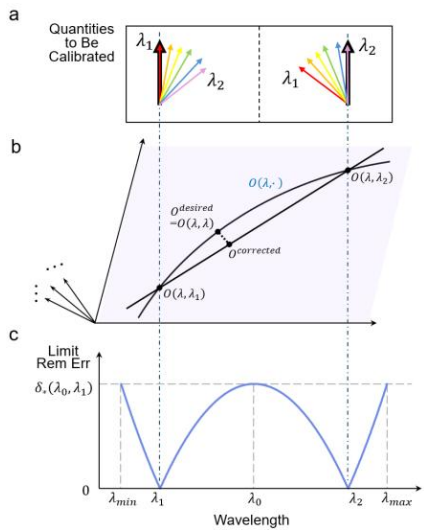
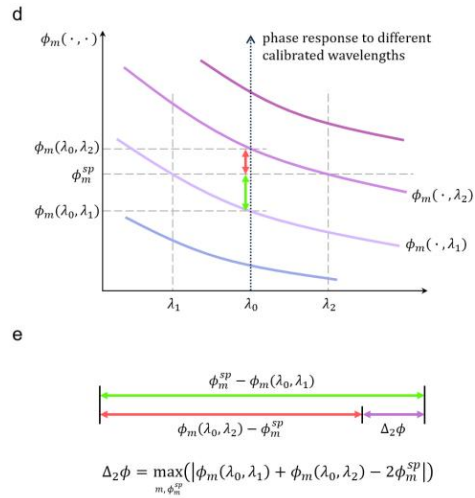

*Contact author: ziqiwei@g.harvard.edu; yuxiao@siom.ac.cn; pengx@siom.ac.cn

†These authors contributed equally to this work

**Fig3. Theory of MZI mesh dispersion error and first order interpolation compensation method.** a) Illustration of multiwavelength optical computing calibration. Target vectors are calibrated at $\lambda_1$ and $\lambda_2$. b) The principle of the first order interpolation compensation method. Take a two-dimensional section ($\cong \mathbb{R}^2$) in $\mathbb{C}^n$ containing the desired output and two measured outputs. The corrected output lies on the interpolation line, with a second-order difference from the desired output. c) The limit remaining error over different wavelengths after applying compensation. It reaches the maximum value $\delta_*(\lambda_0, \lambda_1)$ at the middle and two edges of the computing wavelength range. d) Dispersion behavior of phase shifters can be described by a binary angle function $\beta(\cdot, \cdot)$ over input wavelength and calibrated wavelength, where $\beta$ can be any one of the $n^2$ angles $\theta_m, \phi_m, \alpha_l$. $\beta^{sp}$ is the given specified angle. e) Illustration of the second order dispersion error definition (see Supplementary Information S2 for details).

### B. Practical computation of errors

To bridge theory and experiment, we first analyze single phase shifter which yields angle $\beta$ (can be any one of the $n^2$ angles $\theta_m, \phi_m, \alpha_l$). Assume the effective refractive index of the waveguide $n_{eff}(\lambda, T)$, which is a function over wavelength and temperature. The temperature can be manipulated by applying different voltages on the heater, based on the dependence $T(V)$. The explicit form of $\beta$ is:

$$\beta = 2\pi l \cdot \frac{n_{eff}(\lambda, T) - n_{eff}(\lambda, T_0)}{\lambda} \quad (26)$$

in which $l$ is the heating length, $T_0 = T(0)$ is the local environmental temperature. Assuming the relationship $n_{eff}$ over $T$ is almost linear, the first-order wavelength-dependent variation is approximately:

$$\beta(\lambda, \lambda_0) - \beta^{sp} = \beta^{sp} \frac{\lambda - \lambda_0}{\lambda_0} \cdot b_1 \quad (27)$$

The second-order approximation as follows:

$$\beta(\lambda_0, \lambda_1) + \beta(\lambda_0, 2\lambda_0 - \lambda_1) - 2\beta^{sp} = \beta^{sp} \left(\frac{\lambda_0 - \lambda_1}{\lambda_0}\right)^2 \cdot b_2 \quad (28)$$

The parameters $b_1, b_2$ are the first-order and second-order phase shifter dispersion coefficients (See Supplementary Information S3), respectively. They are only related to the properties of the waveguide materials around the center wavelength $\lambda_0$ and local environmental temperature $T_0$.

When performing matrix calculations, according to Reck and Clements decompositions, the $n^2$ angles $\theta_m, \phi_m, \alpha_l$ the range of values for each of the $n^2$ specified angles $\theta_m^{sp}, \phi_m^{sp}, \alpha_l^{sp}$ is as follows:

$$\theta_m^{sp} \in [0, \pi), \quad \phi_m^{sp} \in [0, 2\pi), \quad \alpha_l^{sp} \in [0, 2\pi) \quad (29)$$

Thus, the upper limit of all the first-order phase deviations using Eq. (10) (27) (29):

$$\delta_1(\lambda, \lambda_0) = \begin{cases} (4n-3)\pi \cdot \left|\frac{\lambda - \lambda_0}{\lambda_0}\right| \cdot |b_1| & \text{triangular} \\ 3n\pi \cdot \left|\frac{\lambda - \lambda_0}{\lambda_0}\right| \cdot |b_1| & \text{rectangular} \end{cases} \quad (30)$$

Similarly, the upper limit of all the second-order phase deviations additionally using Eq. (21) (28):

$$\delta_2(\lambda_0, \lambda_1) = \begin{cases} (4n-3)\pi \cdot \left(\frac{\lambda_0 - \lambda_1}{\lambda_0}\right)^2 \cdot |b_2| & \text{triangular} \\ 3n\pi \cdot \left(\frac{\lambda_0 - \lambda_1}{\lambda_0}\right)^2 \cdot |b_2| & \text{rectangular} \end{cases} \quad (31)$$

Ultimately, the quantitative values of all theoretically discussed errors can be systematically computed, providing a foundation for error correction and system optimization.

## IV. EXPERIMENTAL RESULTS

### A. Experimental setup

The target photonic computing platform implements an MZI network coupled with a microcomb-based multi-wavelength framework. A frequency comb light source centered at 1550nm generates dense spectral lines. These multiplexed optical signals undergo parallel matrix transformations through the reconfigurable MZI network, executing wavelength-parallel matrix multiplications. Following photonic processing, a spectroscope detects the spectrum and the output spectral signatures of such MZI network are extracted.

As an initialization for the experiment, the entire MZI mesh is calibrated at the center wavelength (1550nm) using a monochromatic laser source, which means all $n^2$ angles $\theta_m, \phi_m, \alpha_l$ can be adjusted precisely at the center wavelength.

### B. Observe dispersion-induced error

In order to observe dispersion-induced error, the triangular MZI mesh (as shown in Fig.1a) is configured to establish an optical path that splits light into two waveguides of equal intensity and then converges them. By applying an additional phase to one of two arms (splitting beams), interference cancellation can be achieved at the center wavelength. Owing to dispersion, the light of other wavelengths cannot be eliminated. The remaining error spectrum is mainly dispersion-induced.

*Contact author: ziqiwei@g.harvard.edu; yuxiao@siom.ac.cn; pengx@siom.ac.cn

†These authors contributed equally to this work

In the experiment, a microcomb source is injected into the input port 3, and the spectrum at output port 4 is measured. For simplicity, define the "parallel" state of a single MZI as $\theta = \pi$, the "cross" state as $\theta = 0, 2\pi$ and the "split" state as $\theta = \pi/2$. At first, all MZIs are adjusted to "parallel" and all $\phi's$ equal zero. Then adjust the two MZIs adjacent to the input port 3 and the output port 3 to "50-50-split". The relative phase between two arms can be applied by adjusting the angle $\phi$ of the MZI adjacent to the input port 3, which equals $(2k + 1)\pi$ at interference cancellation, according to the theory.

The remaining error spectra is expected to be mid-split, since lights near the center wavelength are eliminated. Moreover, the remote spectrum is enveloped by a sech$^2$ profile originating from the microcomb. To make the dispersion effect more pronounced, and to suppress other secondary dispersions and wavelength-dependent transmission losses, the optical path difference between the two arms should be above $\sim 5\lambda$, which means $\phi$ needs to be greater than $\sim 10\pi$. Based on these analyses, experiments are conducted and the results are shown in Fig. 4.

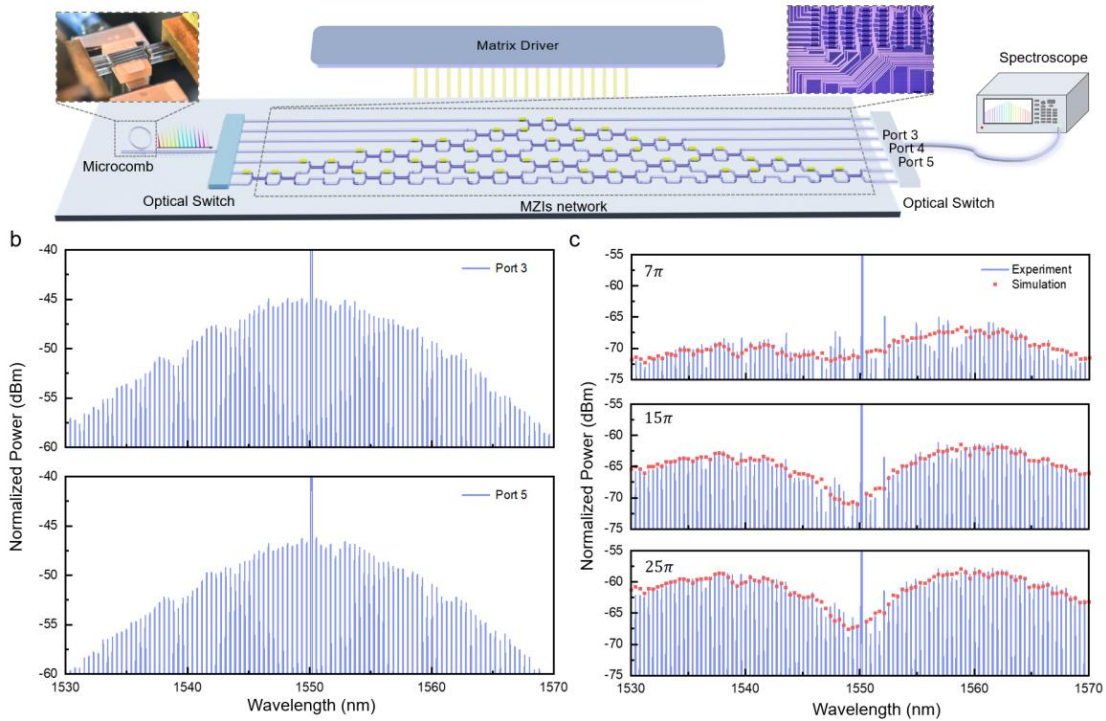

**Fig4. Experimental results on transmission loss and dispersion within MZI mesh.** a) Nearly uniform transmission loss in MZIs is observed by comparing the two spectra. All MZIs have been adjusted to "parallel" or "cross". In the upper diagram, the beam passes through 7 MZIs, while in the lower diagram, it passes through 9 MZIs. b) Two coherent pulsed laser beams of equal intensity interfere in an MZI network. By adjusting the phase shifter in the MZI to achieve extinction at the central wavelength, three spectral profiles were recorded when the optical path difference at the central wavelength was set to $7\pi$, $15\pi$, and $25\pi$. The simulation red dot is based on the dispersion theory mentioned above. The results show side lobes flanking the central wavelength, which are attributed to dispersion effects.

### C. Proof of theoretical prediction

By changing the optical path difference at the central wavelength to $7\pi$, $15\pi$, and $25\pi$, it can be noticed that the remaining error spectra changes significantly (**Fig. 4b**). Specifically, the greater the optical path difference, the higher the remaining error spectra. The mid-split characteristic emerges at $7\pi$, being prominent at $15\pi, 25\pi$; and they are all enveloped by sech$^2$ profile, as predicted. Moreover, the experimental spectra match well with the theoretical simulation spectra (See Supplementary Information S4), further demonstrating the correctness and accuracy of the theory.

*Contact author: ziqiwei@g.harvard.edu; yuxiao@siom.ac.cn; pengx@siom.ac.cn

†These authors contributed equally to this work

For rigor, it is necessary to examine the influence of wavelength-dependent transmission losses experimentally. To implement this argumentation, we adjust the system to the initial state (all MZIs to "parallel" and all $\phi's$ to zero) and inject a microcomb source into the input port 3. First, we measured the spectrum of output port 3. After that, we adjust two MZIs to "cross" so that light of the center wavelength only comes out of output port 5. The output spectrum is shown in **Fig.4b**. In the above two cases, the number of MZIs the beam passes through is 7 and 9, respectively. From these two spectra, we observe that the transmission loss is nearly uniform, which is not the dominant factor of the previous phenomena (**Fig.4c**).

### D. Error correction

In silicon-based waveguide, the approximate values of phase shifter dispersion coefficients $b_1, b_2$ are as follows:

$$b_1 \simeq -1.4, \quad b_2 \sim 10^{-1} \tag{32}$$

which indicates that the deviation $\delta_2(\lambda_0, \lambda_1)$ can be ignored compared to $\delta_1(\lambda_0, \lambda_1)^2$.

Obviously, these mid-split spectra are the first-order dispersion errors, which means the corresponding relative light amplitude difference is proportional to the difference from the center wavelength. According to the experiment, take $\phi$ as $7\pi$, $\lambda$ as 1570nm, the corresponding relative light amplitude difference is -6.6dB, about 0.22 in magnitude.

A numerical verification of the theory was performed by calculating the first-order phase deviation ($\delta_1 = 0.40$ from Eq. (27) with $\beta_{sp} = 7\pi$) and the resulting dispersion error limit (0.22 from Eq. (16)). This computed error limit is equivalent to the experimental relative light amplitude difference of 0.22 in an ideal system, thereby confirming the theoretical predictions. The dispersion error can be reduced by applying the first order interpolation compensation method. After choosing the calibrated wavelengths to be $(1550 \pm 10\sqrt{2})nm$ and performing first order interpolation as shown in Eq. (18), we can eliminate the first-order dispersion error. The uniform limit residual error over the computing wavelength range $[1530,1570]nm$ can be derived using Eq. (24)(27), which equals 0.039, far smaller than the uncompensated value 0.22, confirming the effectiveness of this compensation method. As the compensated uniform limit residual error (second order) is proportional to the square of the uncompensated limit dispersion error (first-order), the smaller the uncompensated dispersion error, the more pronounced the effectiveness of this compensation method.

### V. DISCUSSION

In this work, we developed a comprehensive theoretical framework for parallel optical computing systems based on cascaded MZI mesh, specifically addressing their critical spectral behavior. This theory extends the physical layer modeling of MZI-based optical computing into the multi-wavelength regime. By incorporating the dispersion characteristics of fundamental MZI components, we derived a wavelength-dependent transfer matrix model for the system.

We rigorously analyzed the dispersion-induced error inherent in this matrix, establishing a formulation for the maximal error as a function of both matrix size and operational wavelength range. Experimental results demonstrating phase-related spectral dispersion validate the theoretical model.

Building upon this theoretical foundation, we propose a computationally efficient error correction strategy in spectrum view of point. This method leverages calibration fields measured at specific wavelength channels to establish overall first order correction on intermediate spectral channels. Preliminary estimations indicate that applying a first-order correction significantly reduces the global dispersion error—within a 40 nm range—from 0.22 to 0.039. This strategy achieves effective dispersion mitigation with minimal computational overhead, avoiding substantial consumption of additional resources.

As a foundational contribution to future parallel optical computing chips and algorithms, the theory enables the analysis, calibration, and optimization of fidelity and consistency in broadband MZI parallel computing systems [7]. Furthermore, the analytical insights into the wavelength-dependent behavior of cascaded MZIs can guide the design of more spectrally consistent parallel optical computing chip and open new avenues for ensemble training algorithms. These advances pave the way for significantly broader functionality of parallel optical computing in artificial intelligence, optimization problem solving, and image processing.

### ACKNOWLEDGMENTS

*Contact author: ziqiwei@g.harvard.edu; yuxiao@siom.ac.cn; pengx@siom.ac.cn

†These authors contributed equally to this work

This work is supported by funding from Shanghai Institute of Optics and Fine Mechanics, Chinese Academy of Science (24JR521001) and Shanghai Magnolia Talent Plan Pujiang Project (24PJD125).

## AUTHOR CONTRIBUTIONS

Ziqi Wei, Xiao Yu and Peng Xie conceived the concept. Ziqi Wei, Xiao Yu and Yuanjian Wan developed theoretical model and algorithm. All authors contributed to experiments, data analysis and discussion. All authors prepared the manuscript. Peng Xie supervised the whole project.

*Contact author: ziqiwei@g.harvard.edu; yuxiao@siom.ac.cn; pengx@siom.ac.cn

†These authors contributed equally to this work